\documentclass[usenatbib]{mn2e} 
\usepackage{graphicx}
\DeclareGraphicsExtensions{.pdf,.png,.jpg}

\newcommand{\chandra}{\textit{Chandra}}

\newcommand{\xmm}{\textit{XMM-Newton}}

\newcommand{\atomdb}{{\sc atomdb}}

\newcommand{\zpup}{$\zeta$~Pup}
\newcommand{\xper}{$\xi$~Per}

\newcommand{\taustar}{\ensuremath{\tau_{\ast}}}

\newcommand{\No}{\ensuremath{{N_{\mathrm o}}}}
\newcommand{\Ra}{\ensuremath{{R_{\mathrm a}}}}

\newcommand{\DT}{\ensuremath{{{\Delta}T_{\mathrm \ell}}}}
\newcommand{\Rstar}{\ensuremath{{R_{\ast}}}}

\newcommand{\Rsun}{\ensuremath{\mathrm {R_{\sun}}}}

\newcommand{\Msunyr}{\ensuremath{{\mathrm {M_{\sun}~{\mathrm yr^{-1}}}}}}

\newcommand{\vinf}{\ensuremath{v_{\infty}}}

\newcommand{\Ts}{T_{\rm s}}

\newcommand{\Lya}{${\rm Ly}{\alpha}$}

\newcommand{\Mdot}{\ensuremath{\dot{M}}}

\newcommand{\apj}{ApJ}
\newcommand{\apjs}{ApJS}

\newcommand{\aap}{A\&A}

\newcommand{\mnras}{MNRAS}
\newcommand{\araa}{ARAA}
\newcommand{\pasp}{PASP}
\newcommand{\physrep}{Phys.\ Rep.}

\newcommand{\beq}{\begin{equation}}
\newcommand{\eeq}{\end{equation}}
\newcommand{\beqa}{\begin{eqnarray}}
\newcommand{\eeqa}{\end{eqnarray}}

\begin{document}

\title[O star wind shock-heating rate]{Measuring the shock-heating
  rate in the winds of O stars using X-ray line spectra}
\author[D.Cohen et al.]{David H.\ Cohen,$^{1}$\thanks{E-mail:
    dcohen1@swarthmore.edu} Zequn Li,$^{1}$ Kenneth G. Gayley,$^{2}$ Stanley P.\ Owocki,$^{3}$ 
\newauthor Jon O. Sundqvist,$^{3,4}$  V\'{e}ronique Petit,$^{3,5}$ Maurice A. Leutenegger$^{6,7}$ \\
  $^{1}$Swarthmore College, Department of Physics and Astronomy, Swarthmore, PA 19081, USA\\
  $^{2}$University of Iowa, Department of Physics and Astronomy, Iowa City, IA 52242, USA \\   
  $^{3}$University of Delaware, Bartol Research Institute, Newark,
  DE 19716, USA \\
  $^{4}$Institut f\"ur Astronomie und Astrophysik der Universit\"at M\"unchen, Scheinerstr.\ 1, 81679 M\"unchen, Germany \\
  $^{5}$Florida Institute of Technology, Department of Physics and Space Sciences, Melbourne, FL 32901, USA \\   
  $^{6}$NASA/Goddard Space Flight Center, Code 662, Greenbelt, MD 20771, USA \\
  $^{7}$CRESST and University of Maryland, Baltimore County, Baltimore, MD 21250, USA \\
}

\maketitle

\label{firstpage}

\begin{abstract}
We present a new method for using measured X-ray emission line  fluxes from O stars to determine the shock-heating rate due to instabilities in their radiation-driven winds. The high densities of these winds means that their embedded shocks quickly cool by local radiative emission, while cooling by expansion should be negligible. Ignoring for simplicity any non-radiative mixing or conductive cooling, the method presented here exploits the idea that the cooling post-shock plasma systematically passes through the temperature characteristic of distinct emission lines in the X-ray spectrum. In this way, the observed flux distribution among these X-ray lines can be used to construct the cumulative probability distribution of shock strengths that a typical wind parcel encounters as it advects through the wind. We apply this new method to \chandra\/ grating spectra from five O  stars with X-ray emission indicative of embedded wind shocks in  effectively single massive stars. The results for all the stars are quite similar: the average wind mass element passes through roughly one  shock that heats it to at least $10^6$ K as it advects through the  wind, and the {\it cumulative} distribution of shock strengths is a strongly decreasing function of temperature, consistent with a  negative power-law of index $n \approx 3$, implying a {\it marginal} distribution of shock strengths that scales as $T^{-4}$, and with hints of an even steeper decline or cut-off above $10^7$ K. 
\end{abstract}


\section{Introduction} \label{sec:intro}

Embedded Wind Shocks (EWS) are the source of the ubiquitous soft X-ray
emission seen in O stars. This is confirmed by the significantly
Doppler-broadened X-ray emission lines observed with \chandra\/ and
\xmm\, \citep{Cassinelli2001,Kahn2001,kco2003}. The EWS are generally thought to be
associated with the line deshadowing instability (LDI) that is
intrinsic to any radiation-driven flow in which spectral lines mediate the transfer of momentum from the radiation field to matter \citep{Milne1926,ls1970,ocr1988}. Indeed, hydrodynamics simulations show numerous shocks and associated clumped structure in O-stars winds \citep{Cooper1994,fpp1997,ro2002}.  Modelling indicates that the extent to which the instability is seeded by photospheric variability and limb darkening \citep{fpp1997,so2013} can have a strong effect on the shock structure and X-ray emission that is produced. Additionally, multidimensional effects \citep{do2003,do2005} could also have a significant effect on the shock heating and X-ray emission that thus far has been explored numerically only in one-dimensional simulations. 

Observed X-ray emission can in principle be used to provide constraints on wind models and the physics of the LDI and the associated EWS. However, the X-ray emission levels from EWS in massive stars are affected both by the impulsive shock heating and by the cooling of the post-shock plasma, which can be through both radiative and non-radiative channels.
The fundamental questions are about the efficiency and physics of  the process heating the wind plasma to X-ray emitting temperatures, so it is the nature of the heating 
that we wish to use X-ray data to constrain. 

When multiple cooling channels, including non-radiative ones, contribute significantly, the various cooling processes as well as the heating processes have to be modelled in order to compare theory with observations.  
In the dense winds of many O stars, however, the cooling by radiation is prompt and occurs locally -- the cooling lengths are short.
This situation presents us with the opportunity to parametrize the heating directly, which is the approach taken below. 

However, a complication for our simple radiatively cooled picture has been recently emphasized by \citet{Owocki2013}.  Such rapidly cooling, radiative shocks are subject to thin shell instabilities which can lead to mixing-related reductions in X-ray emission. Because of possible mixing effects, the shock-heating rates we derive in this paper under the assumption of pure radiative cooling represent lower limits. But to the extent that mixing is negligible, then the fact that radiative cooling is the dominant cooling mechanism enables us to extract shock heating constraints directly from observed X-ray emission line spectra of O stars \citep{Gayley2014} by exploiting the fact that the shock-heated wind plasma cools locally via the emission of X-rays in different lines, each one with a different temperature-dependent emissivity. It is from this X-ray emission line spectrum that we can reconstruct the shock-heating rate and also its temperature dependence. 

The X-ray emitting wind plasma in massive stars is usually assumed to
be well described by the coronal approximation: statistical
equilibrium with collisional ionization from the ground state
balanced by radiative and dielectronic recombination, and collisional
excitation from the ground state balanced by spontaneous emission.
This gives rise to a spectrum dominated by emission lines from
modestly excited states to the ground state of highly ionized metals, with a modest contribution
from bremsstrahlung and recombination continuum emission.  Each
emission line has an emissivity that is a relatively peaked, strong
function of temperature, following the temperature dependence of the
ionization balance and excitation rates.  In this way, each line
probes a relatively narrow range of plasma temperatures.

The instantaneous X-ray luminosity from a coronal plasma is simply
equal to the combined emissivity of all the lines (and continuum
processes) multiplied by the emission measure (EM), which is the volume
integral of the particle number density squared. This particular
dependence arises from the two-body nature of the excitation process
of the emission lines (and of the bremsstrahlung and recombination).  The temperature distribution of the plasma has thus
been traditionally characterized by a differential emission measure
(DEM), $\frac{d{\rm EM}}{dT}$, which can be described by a continuous
function or by a sum of isothermal components, perhaps taken to
approximate a continuous distribution with some structure. While
subject to various data and analysis constraints and ambiguities (see e.g.\ \citealt{Gayley2014}),
techniques exist for determining a `best-fitting' DEM from an observed
spectrum (e.g.\ \citealt{Kaastra1996,Liefke2008}).  However, as noted above, such a plasma
temperature distribution combines both the desired information about
shock-heating rates and distributions with extraneous and often
complex and incomplete information about the cooling history of the
hot plasma.

The EM is problematic for another reason as well. Namely, the density-squared dependence means that a given mass of heated plasma will have
a higher EM and radiate faster if it is confined to a
smaller volume. Therefore, the DEM of post-shock plasma will depend not only on the
heating rate and the cooling rate, but also on the local post-shock density, seemingly
causing free parameters of any model that might be fit to data to
proliferate.  However, a key insight about wind shock X-ray emission
that makes the analysis much simpler is that for radiative shocks, the total X-ray fluence from the plasma heated by a shock of a given strength as it cools back down to the ambient temperature depends on the mass that traverses the shock rather than the EM behind the shock at any given instant \citep{Antokhin2004,Owocki2013,Gayley2014}. 

A related consequence is that the spectrum of X-ray photons emitted by this radiatively cooled plasma from the time it is impulsively heated until it returns to its pre-shock temperature is independent of how rapidly the plasma cools \citep{Antokhin2004,zp2007}. In fact, the luminosity of any particular X-ray emission line is simply given by the energy injected into the plasma by its passage across the shock front multiplied by the {\it fraction} of the total plasma emissivity that is due to that line \citep{Gayley2014} -- a result we derive below in detail (and ultimately show in equation \ref{eq:Lelldef3}).  We note the similarity to the treatment of X-ray emission from cooling flows in galaxy clusters \citep{pf2006}. 

Furthermore, since plasma impulsively heated to a given temperature as
it crosses a shock front will radiate at that temperature and, as time
goes on, at every lower temperature until it is fully cooled, an
emission line that forms at a given temperature will probe shocks of
every higher temperature. And the contribution of a plasma mass parcel
to that line will not depend on the heating and cooling history of the
parcel, aside from the requirement that the parcel was heated to at
least the characteristic temperature of the line. In this way, we can
use an ensemble of X-ray emission lines, each with a different
temperature dependent emissivity, to derive not only an overall shock
heating rate but also to derive the cumulative distribution of shock
temperatures. \citet{Gayley2014} has recently presented a comprehensive formalism for analysing such cumulative initial-temperature distributions for any impulsively heated X-ray emitting plasma within the context of simplified heuristic emissivity functions, and we extend that approach here for detailed emissivities of real lines seen in O-star winds. 

In order to obtain the individual X-ray emission line luminosities from the measured line fluxes, a crucial step is to correct not just for the distance to the star, and for the effects of interstellar attenuation, but also for wind attenuation of the emitted X-rays. At any given instant only a small fraction of the wind is shock-heated, while most of it is relatively cold and not highly ionized and so it can efficiently absorb X-rays. While accounting for this wind absorption can be done via detailed modelling of the
spatial distribution of the X-ray emitting wind plasma and the radiation transport through the absorbing wind simultaneously (e.g.\ \citealt{Herve2012}), the procedure is much more tractable when the wind absorption
correction is done separately for each line and independently of the
modelling of the heating and cooling. Here we use the wind absorption optical depths derived by \citet{Cohen2014} via X-ray line-profile fitting to make this wind absorption correction. 

In this paper we apply the cumulative initial-temperature distribution approach described by \citet{Gayley2014} to the specific case of the dense winds of O stars with radiative wind shocks driving
their EWS X-ray emission. Specifically, we analyse the line-rich  \chandra\/
grating spectra of five OB stars  \citep{Cohen2014} in order to derive the shock-heating rates in their winds. In \S2 we describe quantitatively how the
measured line fluxes are related to the wind shock -eating rate and we also describe the data used in the analysis. In
\S3 we present the results for the programme stars. And in \S4 we discuss
the implications of our results. 

\section{Theory and Data Analysis}
\label{sec:theory}

\begin{figure}
\includegraphics[angle=0,width=85mm]{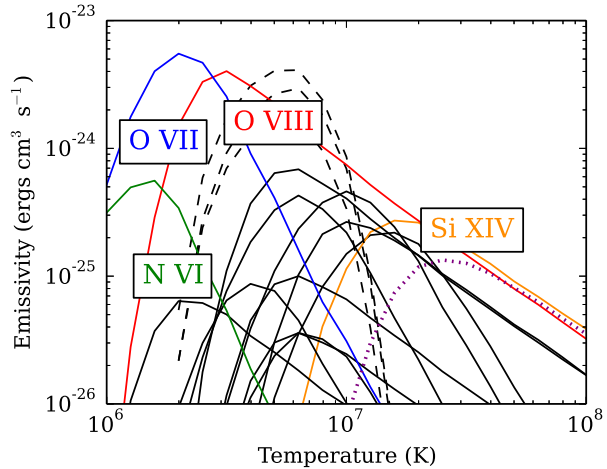}
\caption{The line emissivities from \atomdb\, \citep{Foster2012} for all the lines
  measured in the programme stars. The four lines from each of the
  helium-like complexes are combined together into a single emissivity
  for each complex, and likewise the two components of the \Lya\/ features are combined together into a single emissivity function. Several important lines at the extremes of the
  temperature distribution are labeled and the Fe\, {\sc xvii} lines are denoted by the dashed, black curves. The N\, {\sc vi} line complex is present in \xmm\/ spectra, but not \chandra\/ spectra. We include it in the analysis of \zpup\/ only. Similarly, the dotted purple curve is for the S\, {\sc xvi} \Lya\/ line that we include only in the \zpup\/ analysis (it is not detected, but provides an interesting upper limit, as discussed in sec.\/ \ref{sec:discussion}).  }
\label{fig:emissivities}
\end{figure}

For radiative shocks with negligible mixing all the energy injected into the flow as
it crosses a shock front eventually appears as radiated photons as the
plasma cools from its initial high temperature back down to the
ambient wind temperature. While the plasma remains hotter than about
$10^6$ K most of those photons will be X-rays, whereas at
lower temperatures, most of the radiation will be in the EUV, FUV, and
UV, and will therefore not be observable by \chandra\/ or other X-ray
telescopes. For coronal plasmas the emission strength of a given line, $\ell$, can be
characterized by a temperature-dependent emissivity, $\Lambda_{\ell}(T)$. The form of the
temperature dependence arises from the dependence of the ionization
balance on temperature and, to a lesser extent, from the excitation
rate's temperature dependence. In Fig.\ \ref{fig:emissivities} we show
the emissivities of the lines and line complexes we analyse in the
O star X-ray spectra discussed in this paper. These lines
span more than an order of magnitude in temperature, but with a fair
amount of overlap.

\begin{figure}
\includegraphics[angle=0,angle=0,width=85mm]{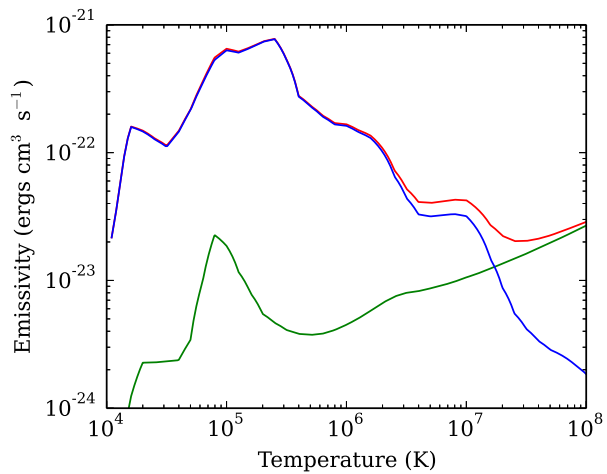}

\caption{The contribution of all emission lines (blue) to the total
  radiated power (red), along with the contribution of continuum
  processes (green).
}
\label{fig:total_power}
\end{figure}

There are many additional weaker lines that contribute to \chandra\/
spectra along with continuum processes -- bremsstrahlung and
recombination -- which are relatively weak for plasmas with
temperatures below 10--20 million K.  In Fig.\ \ref{fig:total_power}
we show the total line-plus-continuum emissivity for a coronal plasma, $\Lambda(T)$,
according to \atomdb\/ \citep{Foster2012}, which is the same source we
use for the individual line emissivities. We note that the \atomdb\/
models assume solar abundances \citep{ag1989}.  More recent
re-evaluations of the solar abundance (e.g.\ \citealt{Asplund2009}),
as well as abundance variations among the specific programme stars,
would lead to factors generally of the order of tens of per cent adjustments to the
line emissivities. We do not account for possible differences in
assumed solar abundances or specific star's particular abundances,
except in a few cases, for which non-solar O and N abundances are quite
significant.  In those cases, we simply scale the \atomdb\/
emissivities for the relevant lines according to the specific element's abundance. Note that
traditionally in X-ray astronomy the quantity referred to as emissivity, $\Lambda$, has units of ergs cm$^3$ s$^{-1}$, so that multiplying it by a number density squared gives the more usual power per unit volume.  The advantage of defining $\Lambda$ in this way is that it is independent of the plasma density. 
 

With these emissivities in hand, we can compute the fraction of the
total radiated power that emerges from the plasma in a given line, at
a given temperature. 
As the hot, post-shock plasma cools back down to its low, pre-shock value, the fraction of radiation emitted in a given line $\ell$ for a single shock with post-shock temperature  $\Ts$, is set by
\begin{equation}
f_\ell (\Ts) = \int_0^{T_{s}} \frac{\Lambda_\ell (T)}{\Lambda(T)} \, \frac{{\rm d}T}{\Ts} 
\, ,
\label{eq:fellts}
\end{equation}
where $\Lambda_\ell (T)$ is the line emissivity (the curves shown in Fig.\ \ref{fig:emissivities}) and $\Lambda(T)$ is the total emissivity [the upper (red) curve shown in Fig.\ \ref{fig:total_power}]\footnote{This approach is similar to the one described in section 2.3 of \citet{Kee2013} for computing the fractional power radiated by a hot plasma into a particular X-ray bandpass.}. Note that it does not matter how quickly the plasma cools, as it will emit the same amount of energy in cooling through some temperature interval whether it does so slowly or rapidly.  It is implicit in our procedure that the stars being analysed have a large, statistical sample of shocks representing all stages of shock evolution and therefore that a single X-ray observation is equivalent to completely tracking the evolution of a representative ensemble of shocks. 

Now, we consider the line emission from a distribution of shocks. Let us suppose a typical fluid parcel undergoes ${\bar N}$ shocks in advecting out through the wind, with each shock having a cumulative probability $p(\Ts)$ for a  post-shock temperature at or above $\Ts$, which declines from unit normalization for very weak shocks, $p(0)=1$.  Then for a wind with mass-loss rate $\Mdot$, the total luminosity in the line is given by an integral over the differential distribution\footnote{The differential probability distribution, $p_{\rm d} \equiv -\frac{d{\rm p}}{d\Ts}$, represents the probability that a shock heats the plasma to a temperature between $\Ts$ and $\Ts + {\delta}\Ts$. Meanwhile, the cumulative probability distribution, $p(\Ts)$, represents the probability that a shock heats the plasma to a temperature of $\Ts$ or lower. For example, if $p_{d}(\Ts)$ is a delta function at $\Ts = T_{\rm o}$ -- representing an impulsive isothermal heating event -- then the cumulative shock distribution function $p(\Ts)$ would be a step function with  $p(\Ts) = 1$ for $\Ts \leq  T_{\rm o}$ and $p(\Ts) = 0$ for $\Ts >  T_{\rm o}$.} in shock temperature, $- \frac{d{\rm p}}{d\Ts}$, multiplied by the energy associated with the shocked wind mass at that temperature and the fraction of the energy that is radiated in the line (given by equation \ref{eq:fellts}),
\begin{equation}
L_{\ell} = \frac{5}{2} \frac{  {\bar N} {\Mdot} k}{\mu m_{\rm p}} \, 
\int_0^\infty - \frac{dp(\Ts)}{d\Ts}  \, \Ts f_\ell (\Ts) \, d\Ts
\,.
\label{eq:Lelldef1}
\end{equation}
Here, the mean molecular weight $\mu$ is in units of the proton mass $m_{\rm p}$ and the post-shock enthalpy per unit mass\footnote{By using the 5/2 enthalpy factor rather than the 3/2 appropriate for internal energy we are allowing for the possibility that the $p{\rm d}V$ work done on the gas as it crosses the shock front could also contribute to the eventual X-ray emission.} is $(5/2) k\Ts/\mu m_{\rm p}$.   Integrating by parts and using  equation (\ref{eq:fellts}) we have
\begin{equation}
L_{\ell} = \frac{5}{2} \frac{ {\bar N}  {\Mdot} k}{\mu m_{\rm p}} \, 
\int_0^\infty \frac{\Lambda_\ell (\Ts)}{\Lambda(\Ts)} p(\Ts) \, d\Ts
\,.
\label{eq:Lelldef2}
\end{equation}

This equation shows that the observed line luminosity, $L_{\ell}$, depends on the convolution of the actual shock distribution function,  $p(\Ts)$, with the emissivity ratio. For notational and conceptual clarity, let us divide this convolution integral into two parts, 
\begin{equation}
\tilde{p}_{\ell}\, \Delta T_{\ell} = 
\int_0^\infty \frac{\Lambda_\ell (\Ts)}{\Lambda(\Ts)} p(\Ts) \, d\Ts,
\label{eq:ptilde}
\end{equation}
where  $\tilde{p}_{\ell}$, the normalized convolution of the cumulative shock distribution, is the observationally derived quantity.\footnote{The quantity $\tilde{p}_{\ell}$ is equivalent to ${\langle p \rangle}_i$ defined in equation (18) of \citet{Gayley2014}.} The  normalization factor, $\Delta T_{\ell}$,  is set by atomic physics,
\begin{equation}
\Delta T_\ell \equiv \int_0^\infty  \frac{\Lambda_\ell (T)}{\Lambda(T)} \, dT,
\label{eq:DeltaTell}
\end{equation}
which represents the portion of  the temperature change in the cooling layer that is associated with a specific line. In an analogy with spectral line equivalent width, this can be thought of as a `temperature equivalent width,'  in that it is the width a line emissivity curve would have if it were rectangular in shape and accounted for all of the radiated power, rather than just a fraction, over that temperature range \citep{Gayley2014}.


\begin{figure}
\includegraphics[angle=0,angle=0,width=41mm]{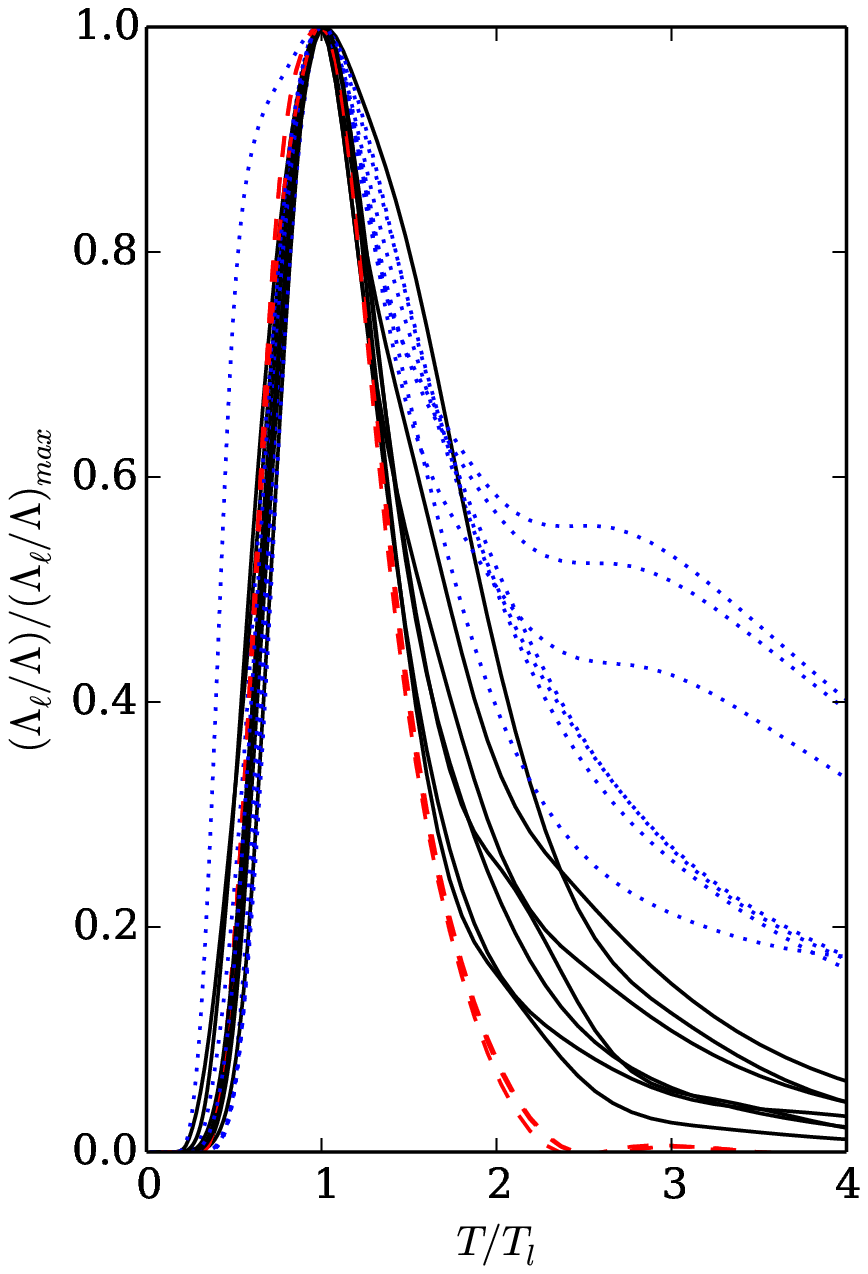}
\includegraphics[angle=0,angle=0,width=41mm]{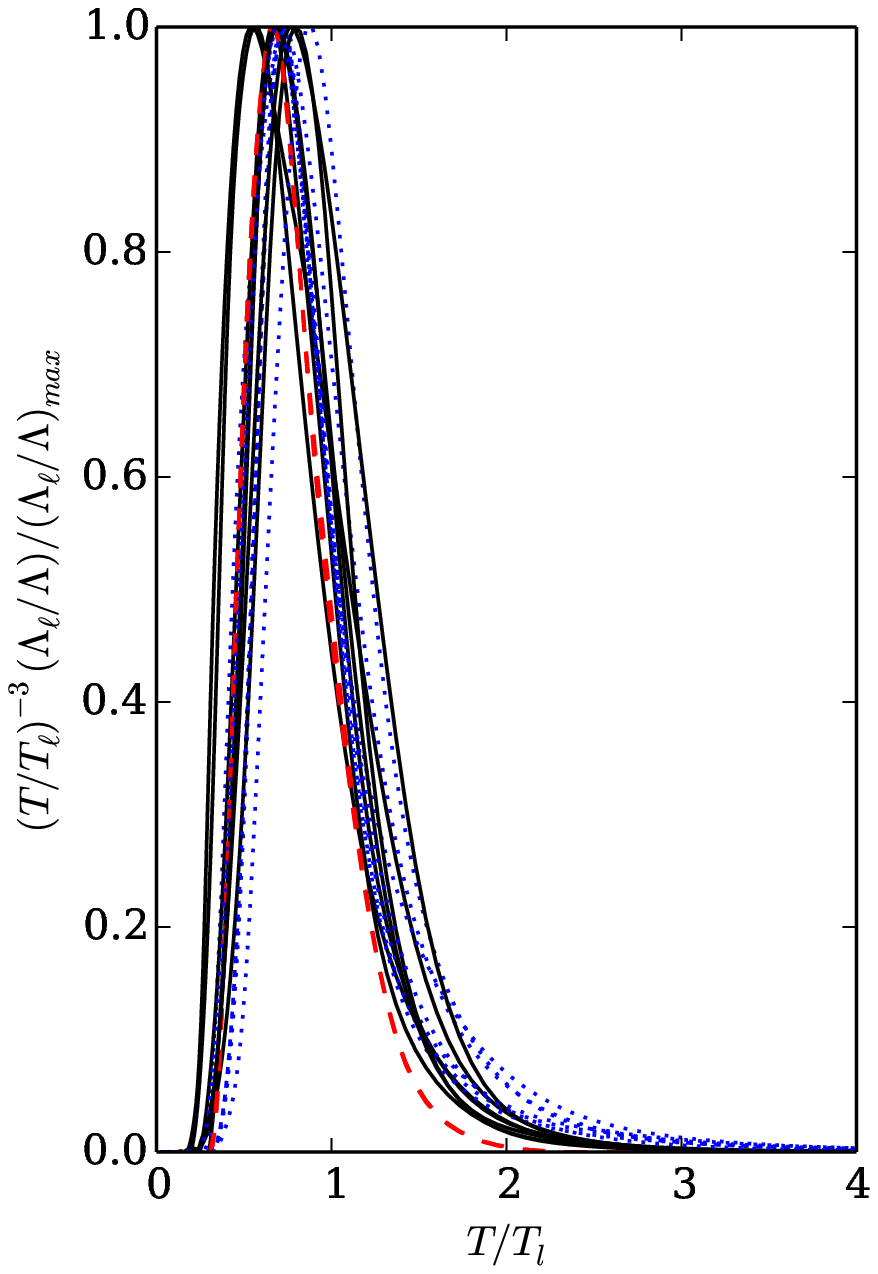}

\caption{The individual line emissivity ratios, $\Lambda_{\ell}/\Lambda$, plotted on a common temperature scale based on each line's $T_{\ell}$, normalized to have the same peak emissivity (left) and also multiplied by a power-law of index $n = 3$ that we find, in the next section, is representative of our programme stars' shock-heating rates, $p(\Ts)$ (right). The quantity in this right-hand panel represents the key integrand in equations (\ref{eq:Lelldef2}) and (\ref{eq:ptilde}). The overlapping red, dashed curves represent iron lines, the solid, black curves, He-like ions, and the dotted, blue curves, H-like ions, with their stronger high-energy tails. Note that the peaks are not at $T = T_{\ell}$ in the right-hand panel because the multiplicative power-law factor affects the emissivity ratios differently depending on their shapes. }
\label{fig:scaled_emissivities}
\end{figure}

Within the assumed model, eqn.\, (\ref{eq:Lelldef2}) is an exact integral expression for the line luminosity given emissivity functions of any form.  As Fig.\ \ref{fig:emissivities} shows, the emissivity for each line has a distinct peak at a specific temperature, and to the extent that these emissivities are strongly peaked functions of temperature, the extraction of the shock-heating rate from an ensemble of line luminosity measurements is quite straightforward, as we will presently show. In Fig.\ \ref{fig:scaled_emissivities}, the individual line emissivity functions are plotted as a ratio of the total emissivity, normalized to have the same peak value, and shown on a temperature scale based on each function's maximum (defined as $T = T_{\ell}$). We also show these functions multiplied by a power-law of index $n = 3$ to approximate the integrand in equation (\ref{eq:ptilde}), which is the key quantity related to the observationally derived shock-heating rate for each line, $\tilde{p}_{\ell}$. Given that this key functional form is quite narrow, and for conceptual and mathematical simplicity\footnote{In Appendix \ref{sec:PBLappendix} we explore a `PLB' model that accounts more realistically for the broader and asymmetric form of the emissivities. A key result is that there is still a direct connection between the observed $\tilde{p}_\ell$ for each line and the shock distribution evaluated at the peak temperature, $p(T_\ell)$, differing only by a normalization factor of order unity from the $\delta$-function based analysis presented in this section.}, here we treat the emissivity ratio as a $\delta$-function,

\begin{equation}
\frac{\Lambda_\ell (\Ts)}{\Lambda(\Ts)} = \Delta T_\ell \, \delta ( \Ts - T_\ell ),
\label{eq:lamdelfun} 
\end{equation}
from which we trivially find that $\tilde{p}_{\ell} = p(T_{\ell})$. Eqn.\, (\ref{eq:Lelldef2}) then becomes
\begin{equation}
L_{\ell}  = {\Mdot} \frac{5k\DT}{2\mu{m_{\rm p}}} \bar{N} p(T_{\ell}),
\label{eq:Lelldef3}
\end{equation}
from which can be obtained an empirically inferred ${\bar N}p(T_\ell )$ in terms of  an observed  set of line luminosities, $L_\ell$, and the tabulated emissivity functions that contribute to \DT,
\begin{equation}
\bar{N}p(T_{\ell}) = \frac{{2\mu{m_{\rm p}}}L_{\ell}}{{5\Mdot}k\DT}. 
\label{eqn:NpT}
\end{equation}
The product $\bar{N}p(T_{\ell})$ is a unitless number
describing the expectation value of the number of shocks  with temperature $\Ts \geq T_{\ell}$ that a mass parcel traverses as it flows through the
wind. 


\begin{table*}
  \caption{Stellar and wind properties}
\begin{tabular}{cccccccc}
  \hline
  Star & Spectral type & $d$ & \Rstar\/ & \vinf\/ & \Mdot\/ & \Ra\/ & $N_{\rm ISM}$  \\
  & & (pc) & (\Rsun) & (km s$^{-1}$) & (\Msunyr) & (\Rstar) & ($10^{22}$ cm$^{-2}$) \\
  \hline
  9 Sgr & O4 V & 1300$^a$ & 12.4$^{b}$ & 3100 & $3.7^{+1.0}_{-0.9} \times 10^{-7}$ & 24 & 0.22 \\
  \zpup\ & O4 If & 460$^c$ & 18.9$^{d}$ & 2250 & $1.76^{+0.13}_{-0.12} \times 10^{-6}$ & 103 & 0.01 \\
  $\xi$ Per & O7.5 III & 382$^e$ & 14.0$^f$ & 2450 & $2.2^{+0.6}_{-0.5} \times 10^{-7}$ & 16 & 0.11 \\
  $\zeta$ Ori & O9.7 Ib & 226$^e$ &  22.1$^b$ & 1850 & $3.4^{+0.6}_{-0.6} \times 10^{-7}$ & 21  & 0.03 \\
  $\epsilon$ Ori & B0 Ia & 363$^g$ & 32.9$^g$ & 1600 & $6.5^{+1.1}_{-1.5} \times 10^{-7}$ & 31 & 0.03 \\
  \hline
\end{tabular}

{References: $^a$\citet{Tothill2008}; $^b$\citet{Martins2005}; $^c$\citet{Markova2004}; $^d$\citet{Najarro2011}; $^e$\citet{vanLeeuwen2007}; $^f$\citet{Repolust2004}; $^g$\citet{Searle2008}; all terminal velocities from $^h$\citet{Haser1995}, all mass-loss rates from \citet{Cohen2014}, and all ISM column densities from \citet{Fruscione1994}  }

\label{tab:stars}
\end{table*}  

The X-ray spectral data we use to make the determination of $L_{\mathrm \ell}$ for use in equation (\ref{eqn:NpT}) are the line
fluxes measured with the \chandra\/ High Energy Transmission Grating
Spectrometer \citep{Canizares2005}.  We supplement these with the N {\sc vi} line complex measured with \xmm\/ in \zpup\/ \citep{Leutenegger2007} to provide information on the low-temperature end of the $p(\Ts)$ distribution. To convert the
measured line fluxes into luminosities, $L_{\ell}$, we apply corrections for (1)
the inverse square law via the distances to the programme
stars; (2) the transmission of the interstellar medium; and (3) the
transmission of the stellar winds themselves.


\begin{figure}
\includegraphics[angle=0,angle=0,width=85mm]{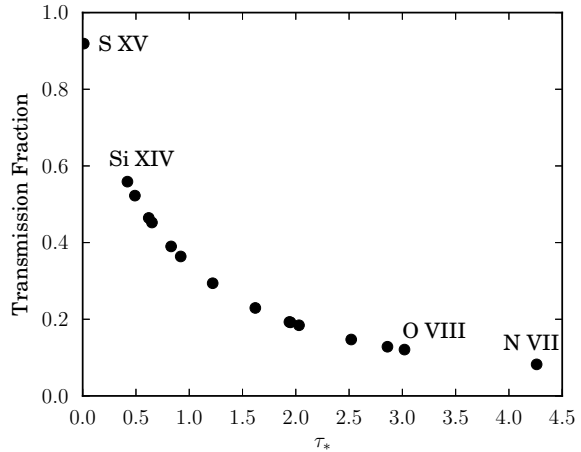}

\caption{The fraction of the emitted line photons that are transmitted
  through the wind without being absorbed, for each line in the
  \chandra\/ spectrum of \zpup, as a function of each line's
  characteristic optical depth value, $\taustar$, derived from fitting
  the line profile shapes \citep{Cohen2014}. }
\label{fig:transmission}
\end{figure}

The bulk, cool component of massive-star winds is a source of
continuum bound-free opacity to the EWS X-rays. Not only does this opacity lead
to attenuation of the X-rays -- which we must correct for --
but it also leads to a characteristic asymmetry of the X-ray line
profiles. The profile fitting of the observed
X-ray emission lines that we use to find the line fluxes also provides
information about the wind optical depth at the wavelength of each
line -- via the parameter, $\taustar
\equiv {\kappa\Mdot}/{4{\rm \pi}\Rstar\vinf}$  \citep{oc2001}. Here $\kappa$ is the wavelength-dependent
bound-free opacity at the wavelength of the emission line, which is assumed to be independent of location in the wind, \Mdot\/ is
the wind mass-loss rate, \Rstar\/ is the stellar radius, and \vinf\/
is the stellar wind terminal velocity.  

From the fitted \taustar\/
values, we can compute the transmission (defined as the fraction of
the emitted X-ray radiation that escapes the wind) using the formalism
of \citet{Leutenegger2010}. We note that this transmission value is
not the usual exponential form due to a slab of absorbing material
between the observer and the source, but rather is a more complicated
function that accounts for the spatial distribution of the emitting
plasma embedded within the absorbing wind.  This wind transmission
correction can be significant. Fig.\ \ref{fig:transmission} shows
the transmission values, $T_{\mathrm w}$, for each of the 16 lines measured in the
\chandra\/ spectrum of \zpup, which is the star in our sample with the
most wind attenuation. Appropriate corrections are applied to the other four stars, as well, with line optical depth values taken from \citet{Cohen2014}. 

Given these considerations, the line luminosity, $L_{\mathrm \ell}$ is
computed from the observed line flux, $F_{\mathrm \ell}$ by

\begin{equation}
L_{\mathrm \ell} = 4{\rm \pi}d^2F_{\mathrm \ell}e^{\tau_{{\rm ism}}}/T_{\rm w}(\taustar),
\label{eqn:LfromF}
\end{equation}

\noindent
where $d$ is the distance to the star, $\tau_{\rm ism}$ is the optical
depth of the interstellar medium, which we compute from the observed
ISM column densities and the {\sc tbabs} ISM absorption model
\citep{Wilms2000}, and $T_w(\taustar)$ is the wind transmission. This is denoted as $T(\taustar)$ in \citet{Leutenegger2010}, but is slightly relabelled here to better distinguish it from the temperature. 

Our sample consists of the five massive stars with high quality \chandra\, grating spectra in \citet{Cohen2014} that are not contaminated by colliding wind X-ray emission and that have winds dense enough to be fully radiative out to large radii (which eliminated the weak-wind star $\zeta$ Oph). We also exclude the O2If* star HD 93129A which has only a small number of lines visible in its \chandra\, spectrum. For each of the five stars analysed, Table \ref{tab:stars} lists the relevant stellar and wind parameters.  This includes the adiabatic radius, \Ra, at which the radiative cooling length of a 1 keV shock equals the stellar radius \citep{Owocki2013}. Below \Ra, shocks cool primarily by radiation, while above it, adiabatic expansion dominates the cooling. The bulk of the X-ray emission from EWS in massive stars comes from the first several stellar radii of the wind according to both theoretical calculations \citep{fpp1997} and observational constraints from line profiles and forbidden-to-intercombination line intensity ratios measured via X-ray spectroscopy \citep{Leutenegger2006}.  Thus the values of \Ra\, listed in the table justify the assumption of radiatively cooled shocks in the programme stars.

\section{Results}
\label{sec:results}

\begin{figure*}

\includegraphics[angle=0,angle=0,width=88mm]{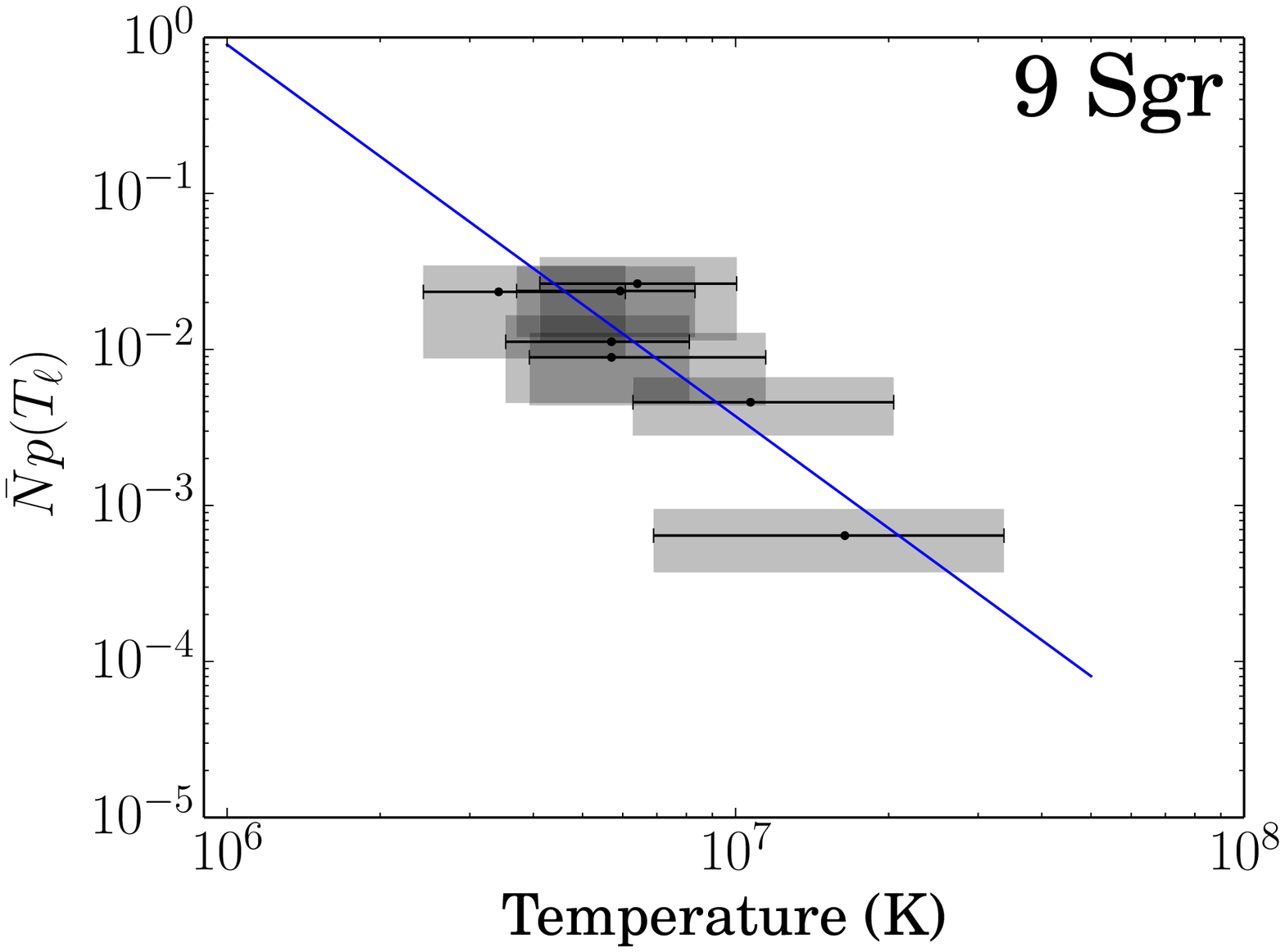}
\includegraphics[angle=0,angle=0,width=88mm]{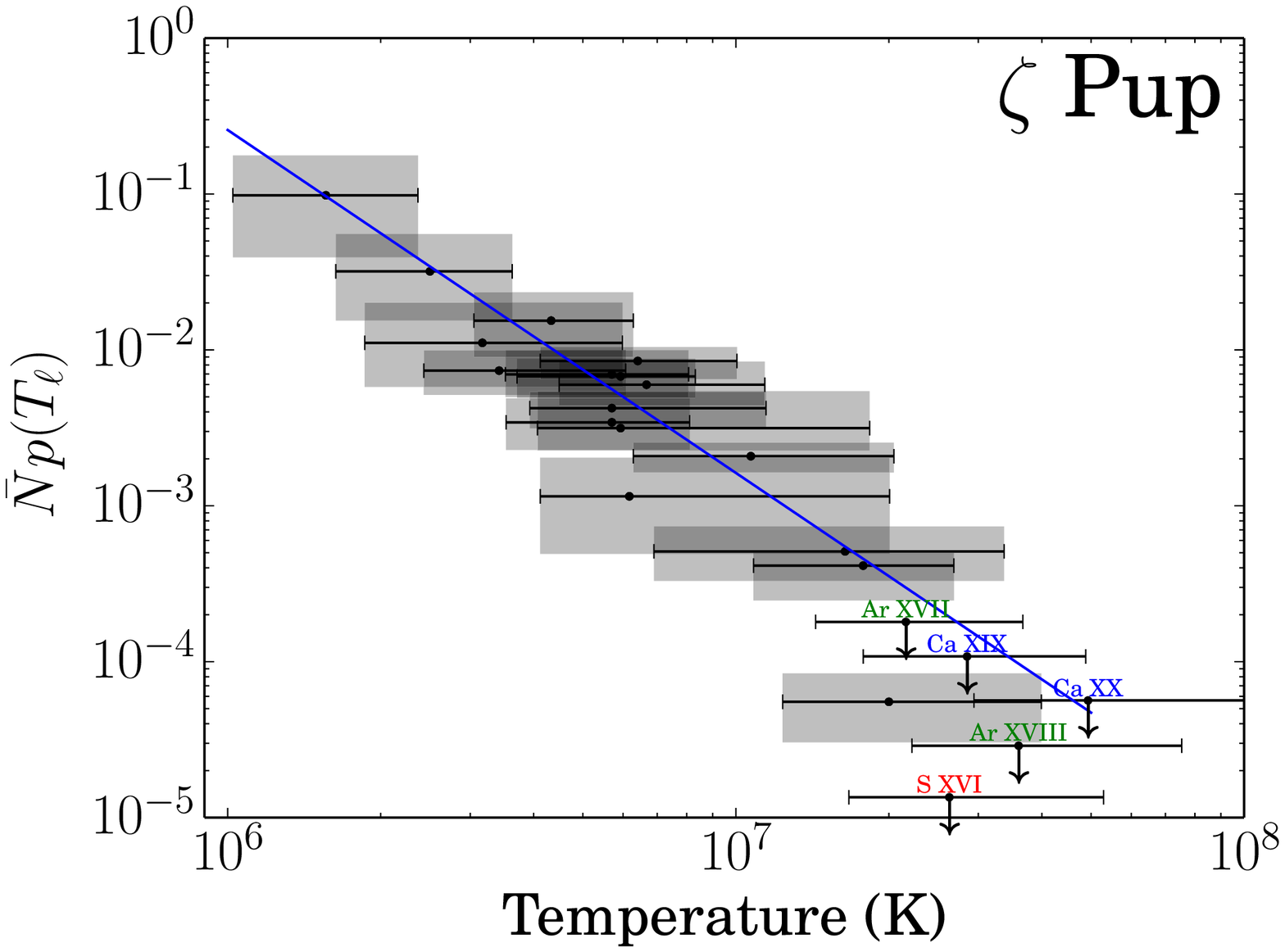}
\includegraphics[angle=0,angle=0,width=88mm]{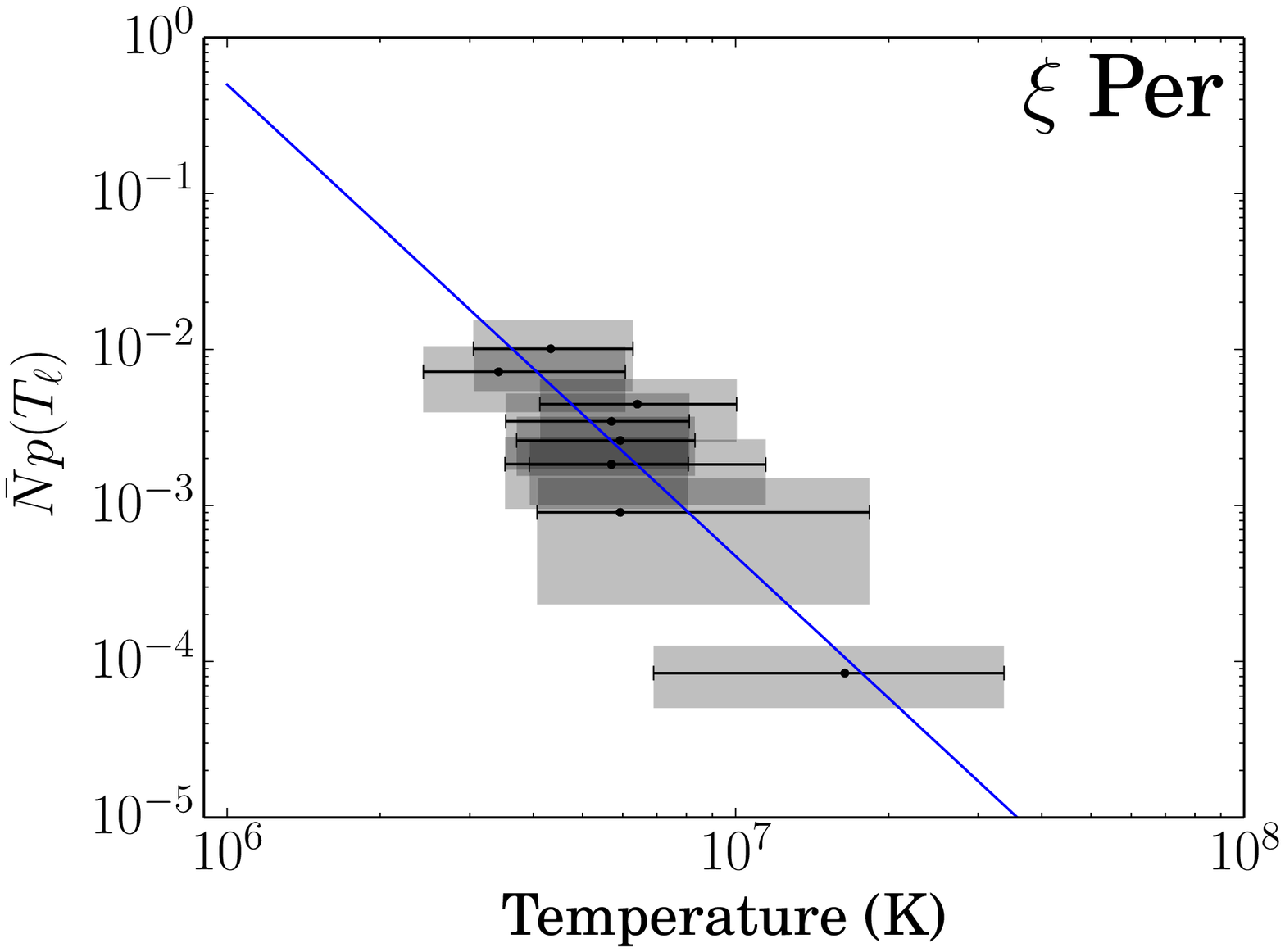}
\includegraphics[angle=0,angle=0,width=88mm]{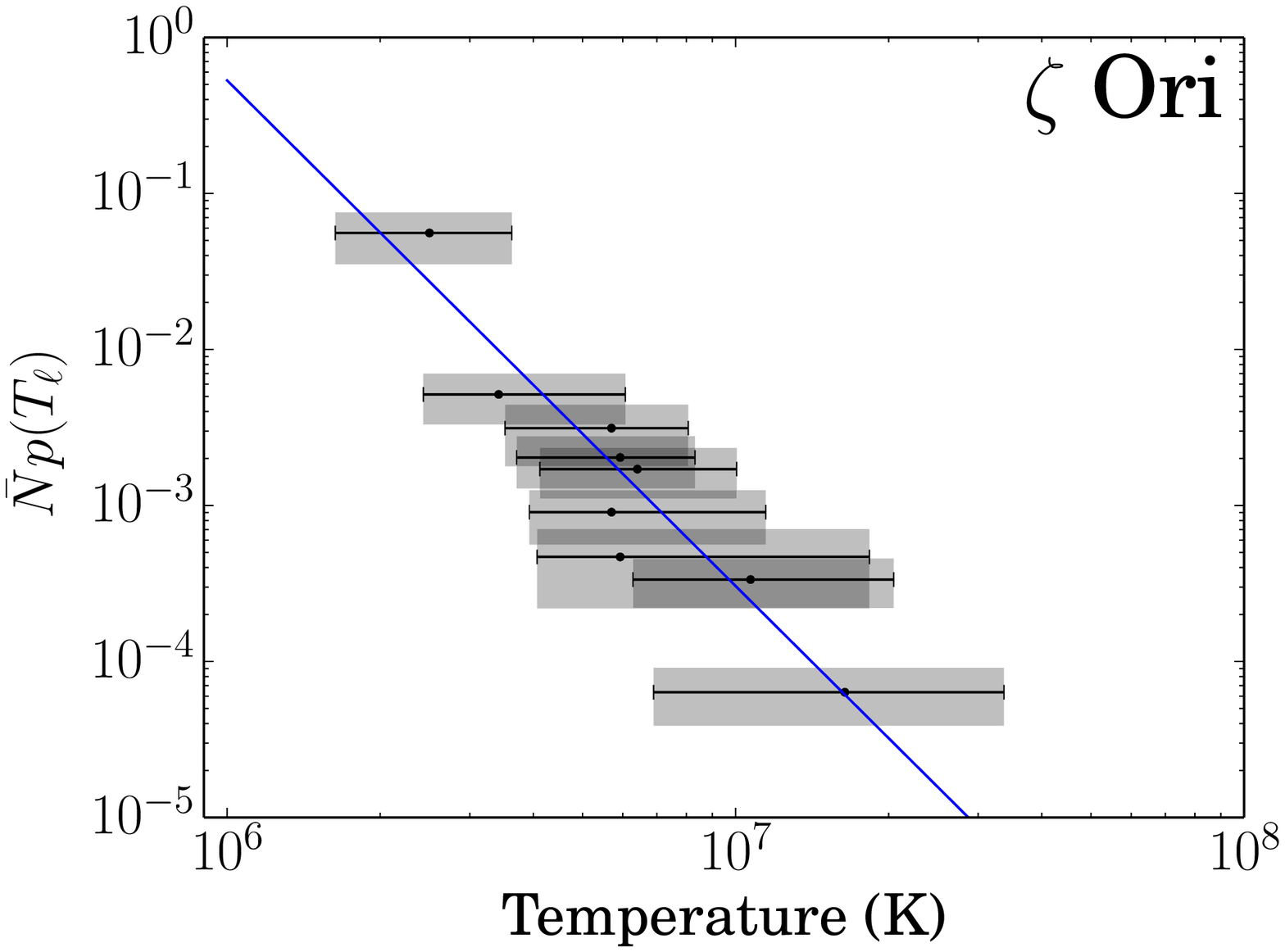}
\includegraphics[angle=0,angle=0,width=88mm]{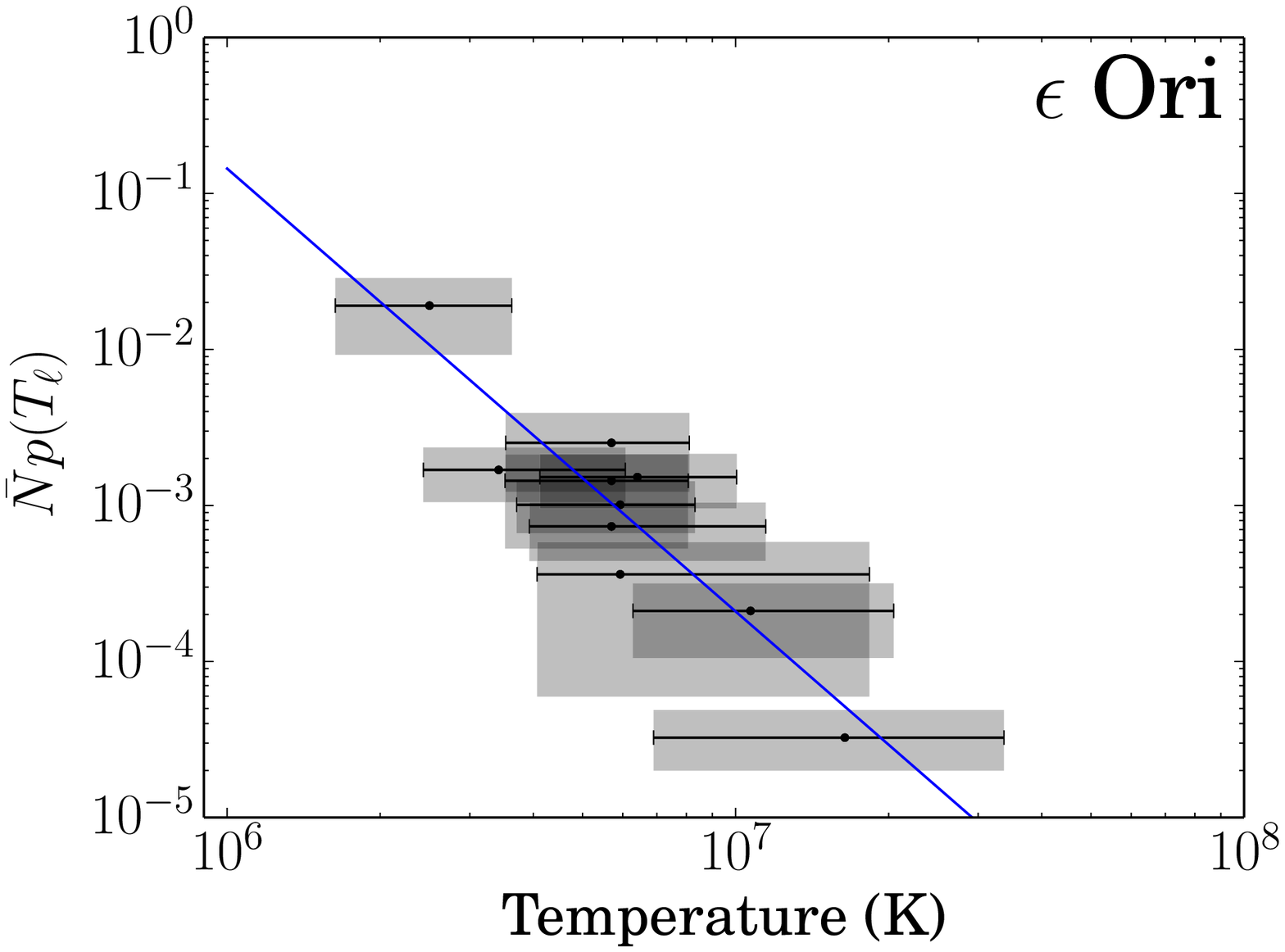}

\caption{The shock-heating rate, $\bar{N}p(T_\ell)$, is shown with the
  uncertainty on its value (vertical extent of each grey box) as well as the FWHM of the line
  emissivity ratio, $\Lambda_{\ell}(T)/\Lambda(T)$ (horizontal extent, visually reinforced by the horizontal error bars). The points are at $T_{\ell}$ for each line. The
  best-fitting power law to each set of values is shown as a blue line in
  each panel. For \zpup, the lowest temperature point corresponds to the N {\sc vi} feature measured with \xmm. And for this star, upper limits are included for five additional lines, none of which are detected in the \chandra\/ spectrum. 
 }

\label{fig:gray_boxes}
\end{figure*}

The shock-heating rates $\bar{N}p(T_\ell)$ computed using equation (\ref{eqn:NpT}) for each
emission line in each of the stars are plotted in Fig.\
\ref{fig:gray_boxes} as a function of the lines' temperatures of peak
emissivity, $T_\ell$. The temperature range for each line in that figure represents the full width at half-maximum (FWHM) of the line emissivity ratio, $\Lambda_{\ell}(T)/\Lambda(T)$.  For each star, different lines that probe similar
temperature ranges give consistent results. And there is clearly a
decreasing trend for each star, consistent with the cumulative,
monotonic nature of the probability function, $p(\Ts)$. Note that this is not something imposed by our method, but rather is a reassuring consistency check on it. We note, also,
that applying the wind and ISM transmission corrections improved the
consistency of the results, as did accounting for lower oxygen and
higher nitrogen abundances in \zpup\/ \citep{Bouret2012,Leutenegger2013} and \xper.\footnote{For this latter star, no nitrogen lines are used, and we make a factor of 1/3 correction to the oxygen abundance, which we base on visual inspection of the X-ray spectrum.} 

Uncertainties in the derived $\bar{N}p(T_\ell)$ values come from several
different sources. The biggest uncertainties are in the stellar and wind properties -- the distances and 
mass-loss rates -- however, those errors will affect every line from a
given star by the same amount and so will simply scale up or down the overall shock
heating rate for a given star. The sources of error that vary from line to line -- statistical error on the measured brightnesses of each line and the ISM and wind transmission corrections -- generally amount to a few tens of per cent. Fig.\ \ref{fig:gray_boxes} shows these latter errors, but not the mass-loss rate and distance errors, as the vertical extent of the grey boxes. 


\begin{figure*}

\includegraphics[angle=0,angle=0,width=155mm]{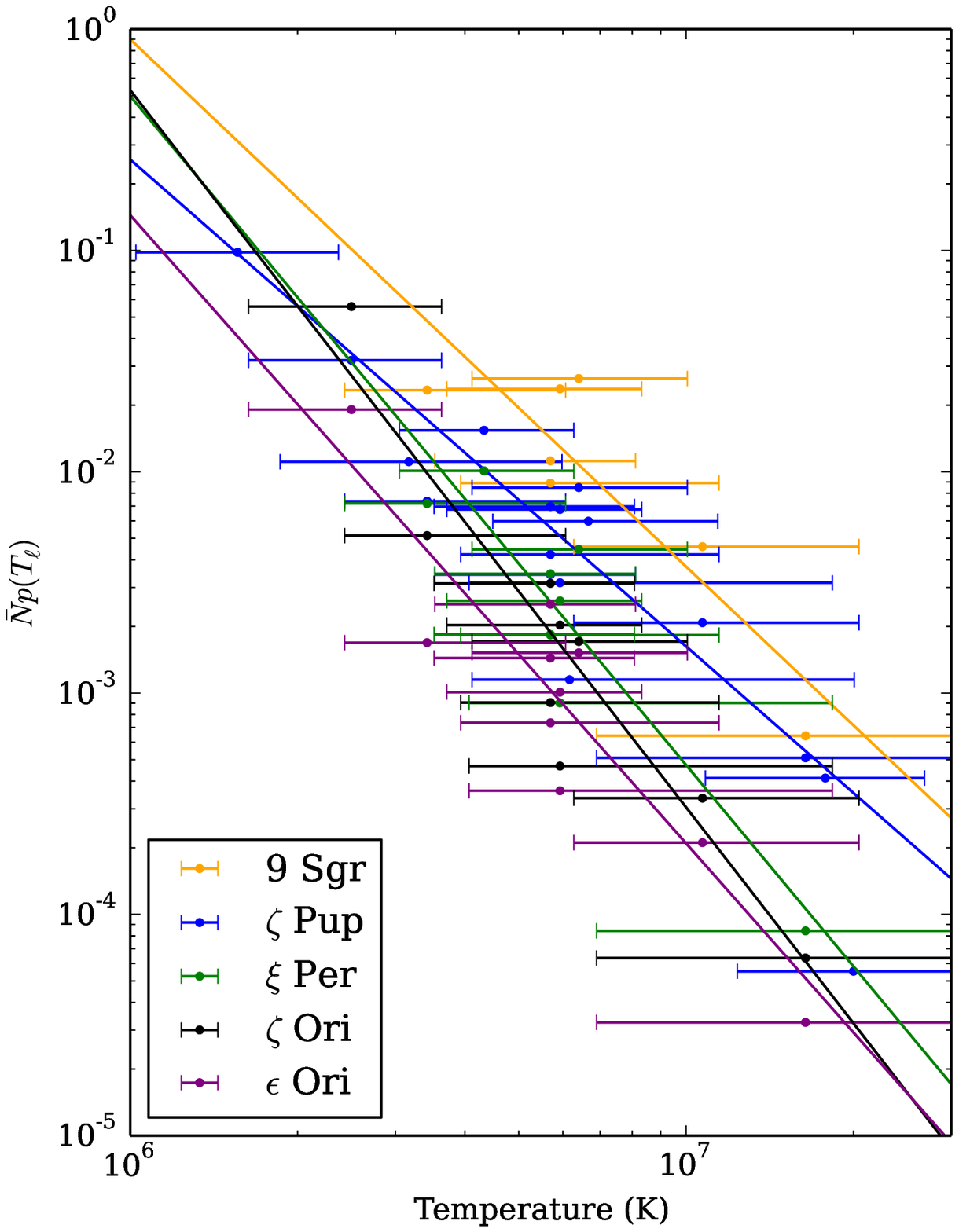}

\caption{The shock-heating rate, $\bar{N}p(T_\ell)$, derived from each line
  of each of our programme stars. These are
  the same results shown in Fig.\ \ref{fig:gray_boxes} but simply
  collected together to facilitate comparison. We do not show the
  uncertainties on each point -- corresponding to the vertical extent
  of the grey boxes -- to keep the plot from being too cluttered.  Because the lower range of the $x$-axis is $T = 10^6$ K, the power-laws' $y$-intercepts give the value of \No\/ for each star, as
  defined in equation (\ref{eqn:powerlaw}).}

\label{fig:NpT_all_stars}
\end{figure*}

We fit a power law to each star's cumulative shock-heating rate -- the ensemble
of $\bar{N}p(T_\ell)$ values -- shown in each panel of
Fig.\, \ref{fig:gray_boxes} and all together in Fig.\, 
\ref{fig:NpT_all_stars}. The power law has the simple form

\begin{equation}
\bar{N}p(\Ts) = N_{\mathrm o}\left(\frac{T}{10^6 ~{\mathrm K}}\right)^{-n}.  
\label{eqn:powerlaw}
\end{equation}

\noindent
Table \ref{tab:powerlaw} lists the best-fitting power-law model parameters, $N_{\mathrm o}$ and
$n$.  All of the measured lines have peak emissivity ratios at temperatures, $T_{\ell} > 10^6$ K, so determining a value for $N_{\mathrm o}$ requires some extrapolation. A less model-dependent statement can be made looking at Fig.\, \ref{fig:NpT_all_stars} --  all the stars have $\bar{N}p(T)$ values approaching 0.1 for their coolest lines, which have peak temperatures, $T_{\ell}$, between 2 and $4 \times 10^{6}$ K. For \zpup, the N {\sc vi} complex observed with \xmm\/ probes temperatures near $1.5 \times 10^{6}$ K and has an $\bar{N}p(T_\ell)$ value slightly above $0.1$.


\begin{table}
  \caption{Power law fits to $\bar{N}p(T_{\ell})$ values}
\begin{tabular}{cccc}
  \hline
  Star & Spectral Type & $N_{\mathrm o}$ & $n$  \\
  & &  &  \\
  \hline
  9 Sgr & O4 V & 0.90 & 2.38 \\
  \zpup\ & O4 If & 0.26 & 2.20 \\
  $\xi$ Per & O7.5 III & 0.50 & 3.02 \\
  $\zeta$ Ori & O9.7 Ib & 0.53 & 3.24 \\
  $\epsilon$ Ori & B0 Ia & 0.14 & 2.84 \\
  \hline
\end{tabular}
\label{tab:powerlaw}
\end{table}  


\section{Discussion}
\label{sec:discussion}

The results for all five stars are quite similar, with the
$\bar{N}p(T_\ell)$ shock-heating rate consistent with a power law index of roughly $n = 3$ and about 10 per cent of the wind mass passing through a shock of $\Ts  > 2 \times 10^6$ K. This temperature is roughly the minimum plasma temperature that produces significant radiation in the X-ray bandpass \citep{ud-Doula2014}, and so our values of $N_{\mathrm o}$ reveal that much of the wind contributes to the observed X-ray emission from these stars, although simulations and observations both show that, at any given instant, only a small fraction of the wind mass is hot enough to emit X-rays. This is, however, consistent with $N_{\mathrm o} \approx 1$ since the typical shock cooling time is much shorter than the wind flow time. 

If we consider the differential probability distribution of shock strengths, rather than the directly derived cumulative one, then we have an even steeper slope of  $T^{-4}$, implying that a shock is $10^4$ times less likely to heat a mass parcel of wind plasma to within 1 K of $T = 10^7$ K than it is to heat it to within 1 K of $T = 10^6$ K. If we consider heating per decade then we are back to having a slope of $T^{-3}$; that is $\frac{d\bar{N}p}{d{\rm log}T} \propto T^{-3}$. So a shock is 1000 times less likely to heat a mass parcel to within a fixed fraction of $T = 10^7$ K as it is to heat it to within that same fixed fraction of $T = 10^6$ K.  

The lines from the hottest plasma tend to be weak, as the strongly decreasing derived shock distributions indicate. We have little direct information about plasma with temperature much in excess of $10^7$ K and, specifically, it is difficult to know whether there is a high-temperature cut-off to the plasma temperature distribution. Bear in mind that the power laws we show in Fig.\, \ref{fig:NpT_all_stars} are fits to the data-derived shock-heating rates for each line, but we have not attempted to show that the distributions truly are power laws. Consulting Fig.\, \ref{fig:emissivities}, we can see that the Si {\sc xiv} line complex near 6 \AA\, does not have significant emissivity much below $10^7$ K, though that line is present in the high signal-to-noise spectrum of \zpup\/ only, while the Mg {\sc xii} \Lya\/ line, with a slightly lower temperature response, is present in each of our programme stars.  So, it does seem safe to say that the shock-heating distribution reliably extends to at least $10^7$ K.  

To further explore the constraints on the hottest plasma and the strongest shocks, for the star \zpup\/ we have extracted upper limits for five high-temperature lines that should be at least moderately strong in coronal plasmas with temperature in excess of $10^7$ K.  We then compute upper limits on the quantity $\bar{N}p(T_{\ell})$ given the measured upper limits for the line luminosities, $L_{\ell}$. These five limits are shown in the \zpup\/ panel in Fig.\ \ref{fig:gray_boxes} as downward facing arrows. The lowest of these, from the S {\sc xvi} \Lya\/ line, probing temperatures between about 15 and 50 million K, is interesting, lying about an order of magnitude below the fitted power-law shock-heating rate, $\bar{N}p(\Ts)$.  Similarly, the shock-heating rate derived from the measured Si {\sc xiv} line, lies substantially below the power-law fit.  As we show in the appendix, for hydrogen-like ions such as these, the empirical shock-heating rate derived for a particular line using equation (\ref{eq:ptilde}), $\tilde{p}_{\ell}$, may underestimate the true shock-heating rate for the line, $p(T_{\ell})$, by about a factor of 2. Correcting for this effect would still leave those two points substantially below the power-law fit, suggesting -- but not providing conclusive evidence -- that there is a strong decline, or even a cut-off, in the shock-heating rate at temperatures above about $10^7$ K. 

The results derived here for the shock-heating rates can be compared to the results from the more traditional DEM approach \citep{ws2005} which finds $\frac{d{\rm EM}}{dT} \sim T^{-2}$ for normal O stars generally (and two of our programme stars, $\zeta$ Pup and $\zeta$ Ori, specifically). The EMs reported on a line-by-line basis in \citet{ws2005} show significantly more scatter than our $\bar{N}p(T_{\ell})$ results do. This is likely due partially to those authors' neglect of wind absorption, which is difficult to account for in the traditional DEM approach. And in the case of $\zeta$ Pup it is also likely due to the neglect of non-solar abundances in the DEM analysis. Finally the steeper slope we find for $\bar{N}p(\Ts)$ can be reconciled with the shallower overall trend \citet{ws2005} report for the O-star DEMs because the DEM should approximate the ratio of the heating rate to the cooling rate, and since radiative cooling in the $10^6 < T < 10^7$ K range is a modestly decreasing function of temperature, the heating rate should indeed have a steeper (negative) slope than the EM.

Finally, we emphasize that our shock-heating rate approach provides a physically meaningful overall normalization, namely the expectation value for the number of shocks a typical wind mass parcel passes through, whereas the DEM provides only a snapshot of the amount of wind material emitting X-rays at any given time. Interpreting that quantity in terms of the physically more informative shock-heating rate would require modelling the cooling as well as the heating, which depends on  the assumptions that are made about the local density in the post-shock X-ray emitting volumes for this density-squared diagnostic. 

Our new results present specific targets for simulations of EWS. They imply an efficient shock heating mechanism, but one that is a strongly decreasing function of shock temperature and which rarely produces shocks hotter than $10^7$ K. And they also strongly suggest that the shock heating mechanism's characteristics are not too sensitive to stellar or wind parameters, as the results for our programme stars are relatively uniform, despite a wide range of spectral subtypes, wind mass-loss rates, and terminal velocities. To the extent that there are differences among the programme stars' results, there appears to be a higher shock-heating rate for the stars with earlier spectral subtypes. Fig.\ \ref{fig:NpT_all_stars} shows that in the middle of the temperature range sampled by the lines we observe, there is roughly an order of magnitude range in the overall shock-heating rate levels among the sample stars. 

It will be interesting to see if the self-excited LDI can reproduce the observational results we have derived here or if perturbations at the wind base and the associated clump--clump collisions \citep{fpp1997,so2013} will be required to explain the results.  
Although there are few relevant predictions in the current literature, numerical simulations of the self-excited instability show velocity dispersions of a few 100 km s$^{-1}$ \citep{ro2002,do2005}, representing how much variation the wind velocity shows at a given radius over a long simulation run-time. This velocity dispersion is largely due to shocks but its magnitude depends on the duty cycle of shocks as well as their strengths, and so it is difficult to interpret directly in terms of a shock-heating rate. The typical velocity dispersion value corresponds to a shock temperature of roughly $10^6$ K $(T_{\mathrm {shock}} \approx 10^6(v_{\mathrm {shock}}/300 {\mathrm {~km~s^{-1}}})^2$ K), consistent with the dominance of weak shocks over strong shocks that we find in this paper. In principle, for one-dimensional numerical simulations especially, it should be possible to track shock fronts and empirically determine both the mass flux and the characteristic shock temperature for each shock and in that way compute the $\bar{N}p(\Ts)$ predicted by theory under various assumptions. We plan to examine this in future work.

\section*{Acknowledgements}

Support for this work was provided by the National Aeronautics and
Space Administration through ADAP grant NNX11AD26G and \chandra\/ grants AR2-13001A and TM3-14001B to Swarthmore College, ATP grant NNX11AF83G to University of Iowa, and ATP grant NNX11AC40G to University of Delaware. JOS was supported by DFG grant Pu117/8-1. ZL was supported by the Physics and Astronomy Department and the Provost's Office of Swarthmore College via a Vanderveld-Cheung Summer Research Fellowship. We thank Randall Smith for his help and advice regarding {\sc atomdb}, and we thank Marc Gagn\'{e} for useful discussions about 9 Sgr.


\appendix 

\section{PLB model for line emission}
\label{sec:PBLappendix}

To infer the wind shock distribution from observed X-ray emission line fluxes, Section \ref{sec:theory} assumes the emissivity ratio $\Lambda_\ell(T)/\Lambda(T)$ can be modelled as a narrow, $\delta$-function form about a peak emission temperature $T_\ell$. But the plot in Fig.\ \ref{fig:emissivities}  suggests instead that this line emissivity has a sharp, exponential, Boltzmann-factor cut-off at low $T$, and a more gradual power-law decline at high $T$. Let us thus examine the effects of such a `Power-Law Boltzmann' (PLB) form for the line emissivity.

Specifically, for the case of a Boltzmann cut-off with an asymptotic  power-index $q>0$ at high temperature, let us write
\beq
\frac{\Lambda_\ell (T)}{\Lambda (T)} \approx 
C_\ell \, e^{- qT_\ell/T}
\, \left ( \frac{T}{ T_\ell} \right )^{-q} 
\, ,
\label{eq:powbolt}
\eeq
which has its peak at temperature $T=T_\ell$, with $C_\ell$ a normalization constant to be set below.
For a shock distribution that is itself a pure power law with index $n$, i.e.\ 
\beq
p(\Ts) = \left ( \frac{\Ts}{T_o} \right)^{-n}
\, ,
\label{eq:powshk}
\eeq
the convolution integral in equation (3) takes the form
\beqa
 \int_0^\infty  \frac{\Lambda_\ell (\Ts)}{\Lambda (\Ts)}  p(\Ts) \, {\rm d}\Ts
 &=&
C_\ell \, T_\ell \, q^{1-n-q} ~ \Gamma ( n+q-1 ) \, \left ( \frac{T_\ell}{T_o} \right  )^{-n} 
\label{eq:convolpow0}
\\ 
&\equiv& 
C_\ell \, T_\ell \, K_{nq} ~ p(T_\ell )  ~~;~~ n+q > 1 \, , ~~
\label{eq:convolpow}
\eeqa
where $\Gamma$ is the complete Gamma function, and the restriction $n+q > 1$ is required to insure convergence of the integral.
In defining a constant factor $K_{nq}$ that depends on the power exponents $n$ and $q$,
the latter equality (\ref{eq:convolpow}) shows that the convolution over this more general PLB form for the emission still yields the original shock distribution at peak emission temperature, $p(T_\ell )$, but now just with a somewhat different normalization.

Note in particular that for $n=0$ and thus $q>1$,  the integral (\ref{eq:convolpow0}) reduces to the definition (5), giving now 
$\Delta T_\ell = C_\ell T_\ell K_{0q}$.
For cases with $q>1$, we can use this to eliminate the normalization constant $C_\ell$ in favour of $\Delta T_\ell$, yielding
\beqa
{\tilde p}_\ell &\equiv& \int_0^\infty  \frac{\Lambda_\ell (\Ts)}{\Lambda (\Ts)} \,  p(\Ts) \, \frac{{\rm d}\Ts}{\Delta T_\ell} 
\\
&=&  \frac{\Gamma(n+q-1)}{q^n \Gamma (q-1)} \, p(T_\ell ) 
\\
&\equiv&  \frac{K_{nq}}{K_{0q}} ~ p(T_\ell ) ~~ ; ~~ q > 1
\, ,
\label{eq:ptilpow}
\eeqa
which thus provides a convenient notation for comparison with results from  \S 2. 

Rather remarkably, we thus find that, for this PLB form for line emission, the emission-weighted integration over the shock distribution, ${\tilde p}_{\ell}$, still reproduces the actual shock distribution, $p(T_\ell )$, with now just an added renormalization factor.

Thus, at least in the case that the original shock distribution is indeed a pure power law, all the previous results -- derived under the assumption that the emission has a narrow, $\delta$-function form about the peak  -- can still be retained for this more realistic PLB form, {\em if} one simply makes a modest renormalization in the overall level of the inferred distribution!
%
%

As a specific example, for He-like ions, like O {\sc vii} or N {\sc vi}, we find $q \approx 3$. For a typical inferred shock distribution power index $n \approx 3$,  this gives
\beq
{\tilde p}_\ell 
 = \frac{8}{9} \, p(T_\ell) ~~ ; ~ n=3,  \, q=3
\, ,
\label{eq:ptn2q3}
\eeq
showing that in this case the renormalization correction is just slightly below unity.
%

For H-like ions like O {\sc viii} or Si {\sc xiv},  we find $q \approx 1.7$, giving,
\beq
{\tilde p}_\ell 
 = 0.65 \, p(T_\ell) ~~ ; ~ n=3,  \, q=1.7
\, ,
\label{eq:ptn2q3b}
\eeq
showing that in this case the renormalization correction is still just moderately below unity.

\end{document}